\documentstyle[a4wide,psfig,epsf,11pt,amsfonts]{article}

\def\Simp#1#2{\left( \begin{array}{c} {#1} \\ {#2} \end{array} \right)}
\def\simp#1#2{\left( {{#1} \atop {#2}} \right) }

\def\div{\mbox{div}}

\def\S0{ \{ S_0 \} }

\def\DC{{\cal D}}
\def\EC{{\cal E}}

\def\NC{{\cal N}}

\def\ZC{{\cal Z}}

\def\fact2#1{#1!^{[2]}}
\def\factn#1#2{#2!^{[#1]}}
\def\McMahon
{W^{3 \rightarrow 2}_{k,l,p} =  \displaystyle{{(k+l+p-1)!^{[2]} (k-1)!^{[2]}
(l-1)!^{[2]} (p-1)!^{[2]}}\over{(k+p-1)!^{[2]} (l+p-1)!^{[2]}
(k+l-1)!^{[2]}}}}

\def\VdM{\mbox{VdM}}

\def\vect#1{\mbox{\boldmath{$#1$}}}

\def\Z{\mbox{\boldmath{$Z$}}}
\def\R{\mbox{\boldmath{$R$}}}

\def\cotan{\mbox{cotan}}

\def\H0#1{H^0_{#1}}

\def\fm{\phi_{\mbox{\scriptsize max}}}

\def\deuxfigs#1#2{
        \begin{center}
        \begin{tabular}{cc}
        \parbox{2in}
        {\begin{center}
        \vfill 

        \ \hspace{0cm}  \epsffile{#1} \

        \vfill 
        \end{center}}
        &  \parbox{2in}
        {\begin{center}
        \vfill 
        
        \ \epsffile{#2} \

        \vfill
        \end{center}} 
        \end{tabular}
        \end{center}
}

\def\ra{\rightarrow}

\title{Entropy and Boundary Conditions in Random Lozenge Tilings}

\author{N. Destainville \medskip \\
{\em Groupe de Physique des Solides, Tour 23-24, 5$^e$ étage,} \\
{\em Universités Paris 7 et 6,} \\
{\em 2, place Jussieu, 75251 Paris Cedex 05, France.}}

\begin{document}

\maketitle

\begin{center}

\parbox{15cm}{
\begin{abstract}
  The tilings of lozenges in 2 dimensions and of rhomboedra in
  3 dimensions are studied when they are constrained by fixed
  boundary conditions. We establish a link between those conditions and
  free or periodic boundary ones: the entropy is written as a
  functional integral which is treated via a saddle-point method. Then
  we can exhibit the dominant states of the statistical ensemble of
  tilings and show that they can display a strong structural
  inhomogeneity caused by the boundary. This inhomogeneity is
  responsible for a difference of entropy between the studied fixed
  boundary tilings and free boundary ones. This method uses a
  representation of tilings by membranes embedded in a
  higher-dimensional hypercubic lattice. It is illustrated in the case
  of 60 degree lozenge tilings.
\end{abstract}}
\end{center}

\bigskip

\noindent {\bf keywords:} Configurational entropy -- Random tilings --
Boundary conditions -- Quasicrystals

\bigskip


\newpage

\section*{Introduction}

Since the discovery of quasicrystals in 1984 \cite{Shechtman84}, a
great deal of work has been accomplished to get a precise microscopic
structural description of these metallic non-crystalline alloys.  It
was rapidly understood that Penrose-like \cite{Penrose74} tilings
could provide very good microscopic models of quasicrystals: it is
highly presumed that favored atomic motifs form tiles. However, the
best description for real quasicrystals remains an open question:
according to the mechanisms involved to describe the structure and
explain its stability, the studied tilings can be perfect Penrose-like
arrangements of tiles or random ones. Indeed, despite their random
character, the latter exhibit global quasiperiodic symmetries
\cite{Elser,Henley91} and are therefore good candidates for
quasicrystal models. The random tilings use the same tiles as the
perfect ones. But in the former, local rearrangements of tiles --
called localized phason flips~-- are allowed. These degrees of freedom
give access to a great number of microscopic configurations. This is
the random tiling model (RTM) \cite{Widom89,Strandburg89,Henley91}. It
involves an important contribution of the tiling entropy to the total
configurational entropy, and therefore to the free energy. This
phenomenon is supposed to favor the quasicrystal against other
competitive phases.

Among the different techniques developed to estimate this tiling
configurational entropy, the partition method
\cite{Elser84,Mosseri93,Mosseri93B,Bibi97,These} presents the
advantage to set a particularly well defined combinatorial problem.
However, the boundary conditions of the tilings considered in this
method are different from the usual ones. As a consequence, the
configurational entropy per tile of partition tilings is lower that
the usual free boundary one: in the simplest case of 60 degree lozenge
tilings, the respective values can be exactly calculated and are about
0.261 \cite{Elser84} and 0.323 \cite{Wannier50} at the infinite size
limit when the 3 fractions of tiles are equal (the configurational
entropy per tile is simply the logarithm of the number of tilings
divided by the number of tiles).

Elser \cite{Elser84} explicited a qualitative argument to explain this
difference. The goal of the present paper is to go further: we will
establish the link between the different kinds of boundary conditions
and we will give a qualitative as well as quantitative explanation for
the difference of entropy per tile.  Hence, we will connect two
related models of statistical mechanics, which are usually
treated via rather unrelated methods.

\medskip

A preliminary part of these results were briefly exposed in reference
\cite{Bibi97}. Most of them were concisely presented in a shortened
version during the Sixth International Conference on Quasicrystals
(ICQ6), in Tokyo \cite{Bibi95}.

\section{Random Lozenge Tilings and Boundary Conditions}
\label{Loz.Til}

In this paper, we will consider $d$-dimensional tilings of rhombic
tiles (lozenges in 2D, rhomboedra in 3D) which tile a region of the
Euclidean space, without gaps nor overlaps. The standard method for
generating such structures consists in a selection of sites and tiles
in a $D$-dimensional lattice ($D>d$) according to certain rules,
followed by a projection onto a suitable $d$-dimensional subspace and
along a generic direction. We then say that we have a $D \ra d$ tiling
problem. It is the so-called ``cut-and-project'' method
\cite{Elser,Duneau,Kalugin}. By construction, the so-obtained rhombic
tiles are the projections of the $d$-dimensional facets of the
$D$-dimensional hypercubic lattice. There are $\simp{D}{d}$ different
species of tiles. In the simple $3\rightarrow 2$ case, this amounts to
three kinds of differently oriented 60 degree lozenges.

Usually, those tilings have periodic or free boundary conditions and
it is generally admitted that the respective entropies are equal at
the thermodynamic limit~-- for given fractions of tiles.  As we have
just seen it, we will consider another kind of boundary
conditions in the following, the fixed boundary ones \cite{Bibi97},
related to the partition method. Then, the region to be tiled will be
the generic ``shadow'' of a $D$-dimensional rectangular
parallelepiped, the sides of which take integer lengths in the
$D$-dimensional hypercubic lattice. This generic shadow is called a
zonotope \cite{Coxeter}, denoted by $\ZC$. When $d=2$, the zonotopes
coincide with the $2D$-gons. The tiles are supposed to perfectly fit
with the boundary $\partial \ZC$ of $\ZC$. Examples of $3 \ra 2$
tilings are given in figures \ref{ex.memb} and \ref{arctic}.

Then the entropy per tile is a function of the different fractions of
tiles, or in other words a function of the side lengths of the
zonotope $\ZC$. For example, the $3 \ra 2$ tilings are enumerated by
MacMahon's formula \cite{MacMahon}, which was derived at the beginning
of this century:
\begin{equation}
\McMahon .
\end{equation}
Here, this formula has been rewritten in terms of second order generalized
factorial functions~\cite{Mosseri93}:
\begin{equation}
\factn{0}{k} = k, \hspace{0.5cm} \factn{m}{k} = \displaystyle{\prod_{j=1}^k
\factn{m-1}{j}}.
\label{gen.fact.def}
\end{equation}
The quantities $k,l$ and $p$ denote the side lengths of the hexagonal
boundary.  The case when all these lengths are equal will be called
{\sl diagonal} in the following, and the corresponding entropies will
be called diagonal entropies.

In appendix \ref{ex.Hex}, we rederive this formula via a purely
combinatorial method which will prove to be useful in the following,
the Gessel-Viennot method \cite{Gessel85,Stembridge90}.

In the following we will be interested in the infinite size limit
entropy of such tilings when the different fractions of tiles are
given. This amounts to making the side lengths of the boundary tending
to infinity with their
relative ratios held fixed: the number of tiles goes to infinity while
the shape (but not the size) of the boundary is kept fixed.

\section{Continuous Limit and Functional Integral}

\subsection{Membrane Representation of Tilings}

This point was already developed in previous publications
\cite{Henley91,Bibi97,These} and is closely related to the
cut-and-project method.  Therefore we will only give a brief
presentation of this method.  The main idea is that a random tiling
can be lifted as a $d$-dimensional non-flat structure embedded in a
$D$-dimensional space.

This structure is a {\sl continuous membrane} made of $d$-dimensional
facets of the $\Z^D$ hypercubic lattice. When this membrane is
projected along the suitable direction, the projections of these
facets are precisely the tiles the tilings are made of
(section~\ref{Loz.Til}); its continuous character guarantees the
absence of gaps in the so-obtained tiling. Such a membrane is said to
be {\sl directed} to emphasize the fact that its projection does not
create any overlap.

\begin{figure}[ht]
\begin{center}
\ \psfig{figure=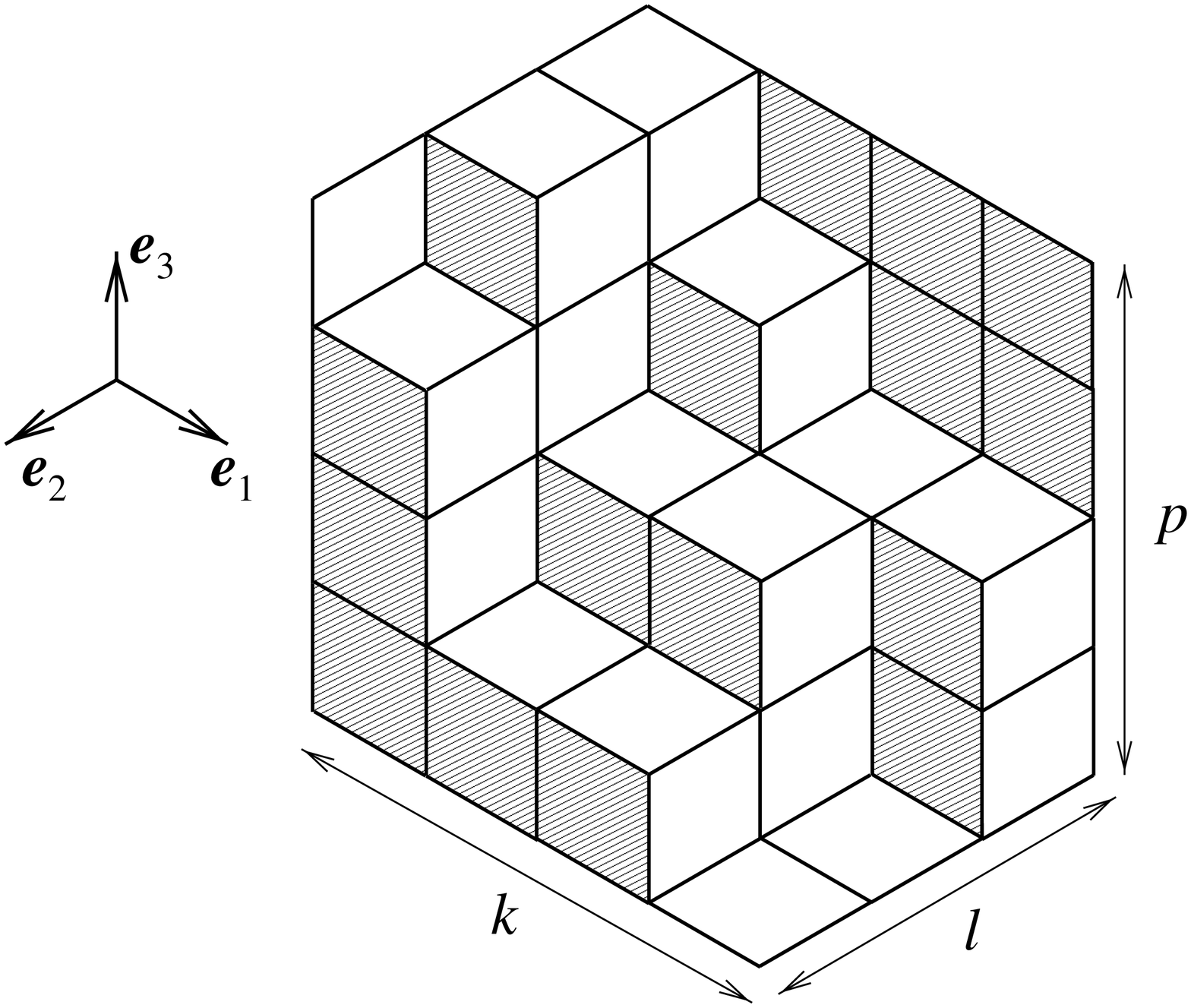,height=5cm} \
\end{center}
\caption{3-dimensional representation of a $3 \ra 2$ tiling.}
\label{ex.memb}
\end{figure}

For example, figure~\ref{ex.memb} displays a $3 \ra 2$ tiling, which
can also be seen as a 2-dimensional non-flat directed membrane
embedded in a cubic lattice. To get a tiling, this membrane must be
projected along the $(1,1,1)$ direction of the cubic lattice. This
point of view can be generalized to arbitrary-dimension tilings. This
correspondence is always one-to-one.

\medskip

In this $D$-dimensional point of view, the $d$-dimensional space on
which the membranes are mapped to get the tilings is called the {\sl
  real space} $\EC$. Its perpendicular space is denoted by
$\EC^\perp$.  For the sake of convenience, we choose the space
coordinates to be the $d$ coordinates on $\EC$ and the $D-d$ 
ones on $\EC^\perp$. Since the membranes are
directed, they can be seen as mono-valuate functions $\phi$ from
$\R^d$ to $\R^{D-d}$. More precisely, in the case of fixed boundary
conditions, these functions are defined on the zonotopal region $\ZC$
of $\EC$.

In the case of free boundary conditions, if the function $\phi$ has a
large scale global gradient $\nabla \phi = \vect{E}$, the random
tiling model states that the fractions of tiles are controlled by this
gradient and, therefore, the entropy per tile can be written as a
function of $\vect{E}$ \cite{Henley91}. This gradient is usually
called the {\sl phason gradient}.

\bigskip

\noindent {\bf Boundary Conditions in the Membrane Representation}

\medskip

We now have to define what the fixed boundary conditions of
section~\ref{Loz.Til} become in the language of directed membranes.
As illustrated on figure~\ref{ex.memb}, we also get a boundary
condition in the $D$-dimensional space: the membrane is inscribed
inside a (non-flat) polygon (or polyhedron), the projection of which
on the $d$-dimensional space gives the tiling boundary, $\partial \ZC$
\cite{Bibi97,These}. For instance, the boundaries of $3 \rightarrow 2$
tilings are (non-flat) $3$-dimensional hexagons. We will call these
boundaries the membrane {\sl frames}. Such a frame can be precisely
characterized as the inverse image, via the projection, of the
boundary of the zonotopal region $\ZC$ which is being tiled
\cite{These}. It is therefore a subset of the boundary of the
hypercube from which $\ZC$ originates. This frame results in
conditions on the functions $\phi$, on the boundary $\partial \ZC$ of
$\ZC$.

\medskip

In the case of free or periodic boundary conditions, the functions
$\phi$ have free or suitable periodic conditions on the boundary of
the domains $\DC$ on which they are defined (here, we use the notation
$\DC$ instead of $\ZC$ to denote domains which might be non-zonotopal).

\subsection{Continuous limit}

Once a tiling has been ``lifted'' in the higher-dimensional space, the
so-obtained directed membrane has a corrugated aspect, due to its
discrete character. Here, we wish to get rid of this discrete
character, at the infinite size limit, in order to study more regular
(``smoother'') objects, to which analytic tools can be applied.
Moreover, these objects will turn out to characterize the macroscopic
states of the statistical ensemble of tilings (or membranes). 
Let us explain how this continuous limit is taken.

\medskip

So far, we have considered tiles of side length 1. To define the
continuous limit, we will get rid of this discrete character, thanks
to a suitable rescaling. For reasons that will become clear in the
following, this side length will go to 0 as the number of tiles tends
to infinity. The functions which represent the tilings are defined on
a domain $\DC$ of the real space. If $N$ is the number of tiles in
$\DC$, we need to rescale the tilings by a factor $1/N^{1/d}$. Thus in
any infinitesimal domain $d^dy$ in $\DC$, the number of tiles goes to
infinity when $N$ does.  Moreover, we do the same rescaling in the
perpendicular space $\EC^{\perp}$.

Once we have done this rescaling, the tilings are represented by
functions $\phi$ which have quite an irregular aspect at small scales.
As it is usually done, for instance in polymer or polymerized membrane
theories, we will treat small scale fluctuations and large scale ones
in a different way. Large scale fluctuations are represented by
regular functions whereas microscopic ones around the latter functions
are integrated in an entropic term. The latter term will have an
exponential form, and will therefore be treated via classical methods
on functional integrals.

Below, we will adopt the following terminology: a membrane
(or function) which has microscopic fluctuations, that is to say which
is the exact representation of a large tiling, will be called {\sl
  faceted}. A membrane, the fluctuations of which have been
integrated in an entropic term, will be called {\sl smooth}. Finally,
we will go on calling a {\sl tile} any $d$-dimensional facet of a
faceted membrane.

\medskip

Given a smooth membrane $\phi$, we can adopt the first naïve
definition of the entropy of this membrane \cite{Bibi97}:
\begin{equation}
s[\phi]=\lim_{N \rightarrow \infty} {\log (\mbox{Number of } N
  \mbox{-tile faceted membranes close to }\phi) \over N}.
\label{s.def}
\end{equation}

The important point here is to understand that, thanks to the previous
rescaling, $\phi$ is kept fixed
while the number of tiles goes to infinity. This point allows, among
many other things, to work with the same set of functions whatever the
system size. 

In this definition, we have not characterized what ``close'' meant.
Let us first state that its precise definition is unessential.  To
understand this point, let us focus on an analogy with a far more
simple statistical mechanics problem, where key ideas are easier to
catch: let us consider a closed box containing a perfect gas. This box
is divided into two parts of same volume, $A$ and $B$, separated by a
virtual frontier. Then one looks for the entropy $\sigma(x_0)$ of this
system, at the thermodynamic limit, when a fraction $x_0$ of the
molecules lie in $A$ (and therefore a fraction $1-x_0$ in $B$). Of
course, this quantity $x_0$ fluctuates and can only be defined up to a
small quantity $\Delta x$. Then we write
\begin{equation}
\sigma(x_0)=\lim_{N \ra \infty} {\log( \mbox{Number of configurations
    such that }x=x_0 \pm \Delta x) \over N},
\end{equation}
where $N$ is the number of molecules. The important point here is that
it can then be proven that this definition of $\sigma(x_0)$ does {\sl
  not} depend, at the thermodynamic limit, on the precise choice of
$\Delta x$, provided it is a finite quantity (see \cite{Toda}, p.~30,
for example, for a discussion on this point).

In particular, if one looks for the more likely value of $x_0$, that
is for the maximum of $\sigma(x_0)$, one finds $x_0=0.5$, still
independently of the choice of $\Delta x$. In other words, the
dominant states are such that $x$ is close to $x_0$, but is not
exactly equal to $x_0$. 

This point of view also applies to directed membranes: we will see in
the next paragraph that this is all the more justified since at the
infinite size limit, only a few constraints cause the faceted
membranes to be ``stuck'' to the smooth one, $\phi$.

\bigskip

The next step consists in considering a point $y_0$ and an
infinitesimal domain $d^dy$ in $\DC$ containing $y_0$. This domain is
large as compared to the tile size, since this size tends to 0.
Moreover, since $d^dy$ is infinitesimal, the gradient $\nabla \phi$
can be considered as constant on this domain. Therefore $d^dy$
contains a piece of tiling with an ``infinite'' number of tiles and a
fixed phason gradient $\vect{E}=\nabla\phi(y_0)$. This phason gradient
is the only constraint on this piece of tiling. In particular, its
boundary conditions are free. Hence if $\sigma(\vect{E})$ denotes the
entropy per tile of a large free boundary tiling of global phason
gradient $\vect{E}$, then the number of faceted membranes close to
$\phi$ and defined on $d^dy$ is equal to
\begin{equation}
\NC(y_0)=\exp \left[ N(d^dy) \sigma(\nabla \phi (y_0)) \right],
\end{equation}
where $N(d^dy)$ is the (large) number of tiles of the previous
membranes. This number of tiles depends on the domain size
(and on the total number of tiles, $N$):
$$N(d^dy)=N. n(\nabla \phi) d^dy,$$
where $n(\nabla \phi)$ is a
function, the integral on $\DC$ of which is equal to 1. Hence
\begin{equation}
\NC(y_0)=\exp \left[ N.n(\nabla \phi (y_0)) \sigma(\nabla \phi (y_0)) d^dy
\right].
\end{equation}

Hence the total number of membranes close to $\phi$ on the whole
domain $\DC$ is given by
\begin{equation}
  \NC_{\phi} = \prod_y \NC(y) = \prod_y \exp \left[ N.n(\nabla \phi (y))
    \sigma(\nabla \phi (y)) d^dy \right].
\end{equation}
This product runs on all the infinitesimal domains $d^dy$. Rigorously,
since the membranes are to coincide on the boundaries between the
different domains, this product should be divided by a boundary
correction term. But when $N \ra \infty$, the latter infinitesimal
domains contain an infinite number of tiles and these boundary terms
disappear\footnote{In other words, the entropy of two infinite size
  sub-systems in which $\nabla \phi$ is fixed is additive.}. The total
number of membranes is therefore equal to the product of the
individual numbers of membranes in each domain $d^dy$ when $N$ is
large.

\bigskip

In order to simplify the forthcoming calculations, we choose the
infinitesimal domains so that they form an hypercubic lattice which
divides the large domain $\DC$ in small cubes of side $a$ and of
volume $a^d=d^dy$. These domains are indexed by Greek letters and
their vertices by Latin letters. 

We have already seen that if $a$ is small enough, then $\phi$ can be
considered as affine on any domain $\alpha$. Therefore this function
only depends on its values $\phi_{\alpha,i}$ on the vertices $i$ of
the domain:
\begin{eqnarray*}
  \NC_{\phi} & = & \prod_{\alpha} \exp \left[ N.n(\vect{E}(\phi_{\alpha,i}))
    \sigma(\vect{E}(\phi_{\alpha,i})) d^dy \right] \\
   & = & \exp \sum_{\alpha} \left[ N.n(\vect{E}(\phi_{\alpha,i}))
    \sigma(\vect{E}(\phi_{\alpha,i})) d^dy \right],
\end{eqnarray*}
which is actually a function $\NC(\phi_i)$ of the values $\phi_i$
on the lattice vertices.

To sum up, we have fixed the global shape of the faceted membranes:
they are tied to the lattice vertices, that is to say we have imposed
their mean gradient on the domains $\alpha$ to be equal to the
gradient of $\phi$. It is actually the {\sl only} constraint we have imposed
to these membranes which are counted by $\NC_\phi$. Now, the key point is
that this constraint is sufficient to be sure that, at the infinite
size limit, the faceted membranes counted by $\NC_\phi$ tend towards
$\phi$. Indeed, Henley \cite{Henley91} showed that for any dimension
$d$, if the gradient $\vect{E}$ is fixed, if $\Delta h(L)$ denotes the
fluctuations in the perpendicular space of directed membranes of
linear size $L$, then
\begin{equation}
{\Delta h(L) \over L} \ra 0 \ \ \ \mbox{when} \ \ \ L \ra \infty.
\end{equation}
(More precisely \cite{Henley91}, if $d=1$, then $\Delta h(L) \sim L^{1/2}$, 
if $d=2$, then $\Delta h(L) \sim \log L$
and if $d=3$, then $\Delta h(L)$ is uniformly bounded). 

After the $1/N^{1/d} \sim 1/L$ rescaling, these fluctuations tend to 0
when $N \ra \infty$. Therefore all the membranes counted by $\NC_\phi$
tend uniformly towards $\phi$. Hence, they are {\sl close} to $\phi$,
whatever the precise definition of this term. Finally,
\begin{equation}
s[\phi] = \lim_{N \ra \infty} {\log(\NC_{\phi}) \over N}.
\end{equation}

Now, the total number of faceted membranes, $\NC$, is given by the
integral\footnote{Rigorously, in the case of free or periodic boundary
  conditions, the problem is invariant under translations of $\phi_i$ and
  this integral $\NC$ is divergent. The membrane must be fixed to a
  point to avoid this divergence. For example, we fix $\phi(y=0)=0$.}:
\begin{equation}
  \NC = \int \prod_i d\phi_i \NC(\phi_i) = \int \prod_i d\phi_i
  \exp \sum_{\alpha} N \left[ n(\vect{E}(\phi_{\alpha,i}))
    \sigma(\vect{E}(\phi_{\alpha,i})) d^dy \right].
\end{equation} 

So far, we have discretized the domain $\DC$ to be sure that the
membranes have an infinite number of tiles in any infinitesimal domain
$\alpha$, at the infinite size limit. Then $\NC_\phi$ or $\NC$ count
faceted membranes close to smooth membranes which are {\sl affine} on
every such domain. To get rid of this restriction, we will now take
the limit $a \ra 0$.  Formally, we write
$$
\DC \phi = \lim_{a \ra 0} \prod_i d\phi_i,
$$
and we turn the sum $\displaystyle{\sum_{\alpha}}$ into an
integral. Thus
\begin{equation}
  \NC  = \int \DC \phi
  \exp \left[ N \int_{\DC} n(\nabla \phi)
    \sigma(\nabla \phi) d^dy \right],
\end{equation}
and 
\begin{equation}
\NC_\phi = \exp \left[ N \int_{\DC} n(\nabla \phi)
    \sigma(\nabla \phi) d^dy \right]. 
\end{equation}
Therefore
\begin{equation}
s[\phi] = \int_{\DC} n(\nabla \phi)
    \sigma(\nabla \phi) d^dy.
\label{s.phi}
\end{equation}
As foreseen, this expression is independent of the precise
characterization of the above ``closeness'' in the definition of
$s[\phi]$. 

\bigskip

This way of writing the entropy associated with a smooth membrane $\phi$
is quite similar to Henley's \cite{Henley91} (section 6.1). He gets
this result thanks to a suitable coarse-graining of the membranes. The
coarse-graining of a faceted function is essentially its local mean
in a neighborhood of diameter $a_0$ of every point of
$\DC$(\footnote{More precisely, the coarse-graining is a convolution
  product of $\phi$ and another function of spatial
  extension~$a_0$.}). Nonetheless, in his point of view, $a_0$ is large but
finite, whereas the smooth functions that we consider here integrate
the local fluctuations of an {\sl infinite} number of tiles when $N$
becomes infinite. Moreover, there is a technical difference in the two
expressions: Henley's entropy is an entropy per unit area, whereas
ours is an entropy per tile. The ratio between these two entropies is
actually $n(\nabla \phi)$, the tile density (per unit area).

Indeed, this last quantity directly depends on the phason gradient
$\vect{E}=\nabla\phi$, since the latter controls the different
fractions of tiles and the different tiles do not necessarily have the
same area. The entropy per unit area, $n(\nabla \phi) \sigma(\nabla
\phi)$, will be denoted by $\tau(\nabla \phi)$.

\medskip

However, in the codimension-one case, all the tiles have the same area
or volume, and equation~\ref{s.phi} can be simplified. The tile
density, $n$, does not depend on the phason gradient any longer and can
be factorized before the integral. The exact value of $n$ depends
on the choice of the rescaling: so far, we have only specified its
order of magnitude ($1/N^{1/d}$) but not its exact value. To calculate
$n$, we choose $\phi$ to be zero everywhere. Then $\nabla \phi=0$ and
$\sigma(\nabla \phi)=\sigma_0$. The entropy per tile $s[\phi]$ is also
equal to $\sigma_0$, since this membrane has a vanishing gradient. Hence 
$\sigma_0=n \int_{\DC} \sigma_0 d^dy$ and $n=1/V(\DC)$, the inverse
volume of $\DC$. 

In the codimension-one case,
\begin{equation}
s[\phi]={\displaystyle{\int_{\DC} \sigma(\nabla \phi) d^dy} \over V(\DC)},
\end{equation}
which is the expression we had given in reference \cite{Bibi97}.

\medskip

\begin{figure}[ht]
\begin{center}
\ \psfig{figure=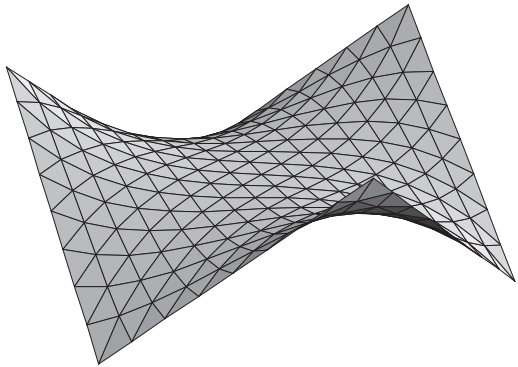,width=7.5cm} \
\end{center}
\vspace{-3mm}
\caption{A smooth function $\phi: \R^2 \ra \R$ 
  and its frame, in the $3 \ra 2$ case.}
\label{ex.phi}
\end{figure}

To sum up, we have coded the macroscopic states of this statistical
ensemble by an internal parameter $\phi$, and we have calculated the
entropy associated with these states. Let us emphasize that the
functional $s[\phi]$ is expressed in terms of the {\sl free} boundary
tiling entropy $\sigma$, whatever the conditions on the boundary
$\partial \DC$ of the domain $\DC$. An example of smooth function
$\phi$ is displayed in figure~\ref{ex.phi}.

\subsection{Dominant States in the Statistical Ensemble}

The faceted membranes defined on $\DC$, and that have the good
boundary conditions, are counted by
\begin{equation}
\NC = \int_{\phi \in F} \DC \phi \
  \exp \left[ N s[\phi] \right].
\end{equation}
Let us precise that this functional integral is taken upon the set $F$
of functions which are smooth representations of faceted membranes.

Moreover, the entropy per tile associated with this set of all the
membranes (or tilings) is given by
\begin{equation}
S = \lim_{N \ra \infty} {\log \NC \over N}.
\end{equation}

Now, we suppose that there exists an unique function $\fm$ that
maximizes the entropy functional $s[\phi]$. This point will be
discussed at the end of this section. Moreover, we suppose that the
functional is non-singular near this maximum, that is to say it has a
quadratic behavior:
\begin{equation}
s[\phi] = s[\fm]- 
\int_{\DC} d^du \int_{\DC}d^dv \ k_{u,v}(\phi(u)-\fm(u),\phi(v)-\fm(v))+
\ldots,
\label{quad.decomp}
\end{equation}
where $k$ is a positive quadratic form, which {\sl a priori} depends
on the point $(u,v)$.

Hence, near $\fm$, $\NC$ is a Gaussian functional integral, and
thanks to a generalized saddle-point argument, we get
\begin{eqnarray*}
S & = & \lim_{N \rightarrow \infty} 
                        {\log(\NC)\over N}   \\ 
  & = & s[\fm]. 
\label{entro.egal}
\end{eqnarray*}
This is a classical result in statistical physics: at the infinite size
limit, the total entropy is equal to the entropy of a dominant
macroscopic state. In other words, the statistical ensemble of
faceted membranes is dominated by membranes close to $\fm$. In the
space of membranes, the distribution is ``peaked'' around $\fm$ at the
infinite size limit, and looks more and more like a Dirac distribution.

\bigskip

To close this section, we must discuss the assumption of unicity of
$\fm$. We will use a general convexity argument: if a function $f$ is
strictly concave on a convex set $C$ and if $f$ has a maximum on $C$,
then this maximum is unique.

Now, the set $F$ of functions is convex: whatever their boundary
conditions on $\partial \DC$, let $\phi_1$ and $\phi_2$ be any two
smooth functions in $F$ and let $\lambda$ be any real number between 0
and 1, then $\phi_\lambda = \lambda \phi_1 + (1 - \lambda) \phi_2$ is
also an element of $F$. In particular, if $\nabla \phi_1$ and $\nabla
\phi_2$ satisfy the good conditions to insure $\phi_1$ and $\phi_2$ to
be in $F$, by linearity of the gradient, $\nabla(\lambda \phi_1 + (1 -
\lambda) \phi_2)$ satisfies the same conditions. And whatever the
boundary conditions, $\phi_\lambda$ also satisfied them.

Let us now check whether $s[\phi]$ is concave: it is an integral, and
therefore a positive linear combination of functions of $\phi$ (the
entropies per unit area). If we prove that all these functions are
concave, then it follows that $s[\phi]$ is concave. Now, the entropy
at the point $y$ in $\DC$ is the composition of $\tau(\vect{E})$ and
the function $\phi \mapsto \nabla \phi(y)$, which is in turn a linear
function. Hence we only need to prove that $\tau(\vect{E})$ is a
concave function of $\vect{E}$. 

This property is a little stronger than the general {\sl random tiling
  model} hypothesis, which states that the free entropy has a
unique maximum as a function of the gradient $\vect{E}$, and is
quadratic near this maximum \cite{Henley91}:
\begin{equation}
\tau^{free}(\vect{E}) \simeq \tau^{free}_0 - {1 \over 2} K^{free} 
\vect{E}^2,
\label{approx}
\end{equation}
where $K^{free}$ is the so-called tensor of phason elastic constants.
Even if our stronger hypothesis is {\sl a priori} valid for a more
restricted set of models, let us note that it is satisfied in all
exactly solvable tiling models \cite{Blote82,Kalugin94,Nienhuis96}.
The concavity of $s[\phi]$ and the unicity of $\fm$ are therefore
reasonable hypotheses.

Finally, let us remark that since this maximum is unique, it will
respect all the underlying symmetries of the problem. We will see an
illustration of this fact in the following.

\section{Relationship Between Different Boundary Conditions}

In principle, whenever the free boundary entropy $\tau$ is known, the
functional $s[\phi]$ is precisely defined, and one can therefore get
the fixed boundary entropy. Theoretically, we are able de deduce the
maximum entropy of fixed boundary tilings, $\tau_0^{fixed}$, as well
as the phason elastic constants, $K^{fixed}$, from their counterpart
in the free boundary case, $\tau_0^{free}$ et $K^{free}$, and to
invert these relations. This was done in the $3 \ra 2$ case, in
reference \cite{Bibi97}(\footnote{In this reference, the value 0.253
  of the diagonal entropy in the quadratic approximation was
  erroneous, because of badly controlled boundary effects. The actual
  value is 0.251.}), in the so-called quadratic approximation, which
consists in estimating the free boundary entropy by its quadratic
development (equation \ref{approx}) near its maximum. Here, we will
go further and give a complete treatment of this $3 \ra 2$ case. We
will also present some related work in the case of different fixed
boundary conditions.

\medskip

To go beyond the quadratic approximation, we will characterize the
maximum of the entropy functional $s[\phi]$ by means of a
functional derivation. If $s[\phi]=\int_{\DC} n(\nabla \phi)
    \sigma(\nabla \phi) d^dy$, then 
\begin{eqnarray}
\delta s & = & s[\phi + \delta\phi ] - s[\phi] \nonumber \\ 
 & = & {1 \over V(\DC)} \int_\DC \vect{d
  \tau}(\nabla \phi(y)) \vect{.\nabla}(\delta \phi) d^dy \\
 & = & {- 1 \over V(\DC)} \int_\DC \delta \phi \ \div(\vect{d \tau})
 d^dy. \nonumber 
\end{eqnarray}
Hence,
\begin{equation}
{\delta s \over \delta \phi (y)} = -\div(\vect{d
  \tau}(\nabla\phi(y))).
\label{deriv}
\end{equation}
Therefore $\fm$ is the function $\phi$ which satisfies this equation
and which has the good boundary conditions.

\subsection{Hexagonal Tilings}

Expression \ref{deriv} is general. In the $3 \ra 2$ case,
it reads
\begin{equation}
{\partial^2 \tau \over \partial E_1^2}
{\partial^2 \phi \over \partial x^2}
+2 {\partial^2 \tau \over \partial E_1 \partial E_2}
{\partial^2 \phi \over \partial x \partial y}
+{\partial^2 \tau \over \partial E_2^2 }
{\partial^2 \phi \over \partial y^2} =0,
\label{edp}
\end{equation}
where $\vect{E}=(E_1,E_2)$.

The coefficients $\displaystyle{\partial^2 \tau \over \partial
  E_1^2}$, $\displaystyle{\partial^2 \tau \over \partial E_1 \partial
  E_2}$ and $\displaystyle{\partial^2 \tau \over \partial E_2^2}$ are
known. Indeed, in the $3 \ra 2$ case, there exists an analogy between
the tilings and the ground states of an antiferromagnetic Ising model
on a triangular lattice \cite{Blote82}. The entropy can then be
derived from the previous solution of this Ising model 
\cite{Houtappel50}. In the latter reference, the entropy is written in
terms of chemical potentials.  Some algebraic manipulations are
therefore necessary to write it in terms of $\vect{E}$, and then get
the above coefficients \cite{These}:
\begin{eqnarray}
\partial^2 \tau \over \partial E_1^2 & = & -{\pi \over 9} {1 \over \sin
  \theta} \left( {1 + w^2 \over 1-w^2} - \cos \theta \right), \\
\partial^2 \tau \over \partial E_1 \partial E_2& = &
{\pi \over 3 \sqrt{3}} \  
{w \over \sin \theta} \ {2 - w^2 \over 1 -w^2}, \\
\partial^2 \tau \over \partial E_2^2 & = & -{2 \pi \over 3} \ {w \over 1
  -w^2}\ {1 \over \sin \phi},
\end{eqnarray}
where $\theta=\displaystyle{{\pi \sqrt{2} \over 3}(E_1 + \sqrt{2})}$,
$\phi=\displaystyle{\sqrt{2 \over 3} \pi E_2}$, and
$w=\tan(\theta/2)\cotan(\phi/2)$. 

\medskip

The partial differential equation \ref{edp} can be solved by means of
numerical calculations. The idea is to discretize the domain $\ZC$,
which is a hexagon in this case, and to use an iterative process: at
each step, a function $\phi_k$ is computed. The above coefficients are
calculated in terms of $\phi_k$. Then $\phi_{k+1}$ is the solution of
a linear system which is the discrete version of equation~\ref{edp}.
Then the sought function is the fixed point of this iterative process,
which is reached after about 10 iterations. This method was used in
reference \cite{These} and gave satisfactory numerical results. 

\medskip

However, more recently, we were aware of related works by 
mathematicians which are interested in similar problems. They
are indeed able to exactly compute the function $\fm$ by the mean of
purely combinatoric methods \cite{Cohn98,Propp98}: the idea is to
calculate the number of fixed boundary tilings with a precise
distribution of vertical tiles upon a given horizontal line. The value
of $\fm$ on this horizontal line is then given by the distribution of
vertical tiles which maximizes the number of such tilings at the
infinite size limit.

\begin{figure}[ht]
\begin{center}
\ \psfig{figure=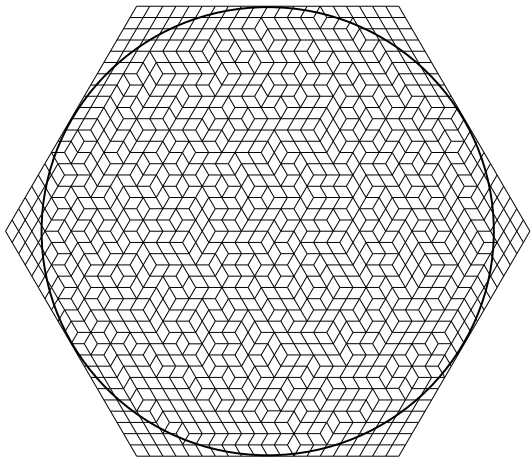,width=7cm} \
\end{center}
\caption{A randomly generated $3 \ra 2$ tiling. At the infinite
  size limit, there is a circular central region (the ``arctic
  circle'' \cite{Cohn98}), where the tiling contains 3 kinds of tiles,
  and 6 ``triangular'' regions, where the tiling contains only one
  kind of tiles and is said to be ``frozen''.}
\label{arctic}
\end{figure}

This exact solution points up a very singular phenomenon. The above
authors called it the {\sl arctic circle phenomenon}: at the infinite
size limit, in the tiling representation in 2 dimensions, there is a
central region of $\EC$ which is circular in the diagonal case and
elliptic in general, inside which the tiling is random~-- in this
sense that it contains the 3 kinds of tiles. Outside this region, the
tiling is ``frozen'': as illustrated in figure~\ref{arctic}, there are
6 regions where there are only one kind of tiles, and where the
entropy is equal to 0.

In these regions, the function $\fm$ is rigorously linear and its
gradient is constant. Inside the central region, in the diagonal case,
the phason gradient of $\fm$ is equal to \cite{Cohn98}:
\begin{eqnarray}
E_1 & = & - \sqrt{2} + {3 \over \pi \sqrt{2}} \left[ \cotan^{-1} f(x,y)
+ \cotan^{-1} f(-x,y) \right], \\
E_2 & = & {1 \over \pi} \sqrt{3 \over 2} \left[ \cotan^{-1} f(x,y)
- \cotan^{-1} f(-x,y) \right],
\end{eqnarray}
where $f(x,y)=\displaystyle{{1 \over 2 \sqrt{3}} \ {{8/\sqrt{3} \ x y
      - 8/3 \ y^2 + 2 } \over \sqrt{1 - 4/3(x^2 + y^2)}}}$, if the
side of the hexagon has been rescaled to 1. In this expression and the
previous ones, the origin of the coordinates is at the center of the
hexagon, the axis $(Ox)$ is pointed towards a vertex of the hexagon.

\medskip

These expressions characterize $\fm$. In fact, the resulting function
is very close to the function showed in figure~\ref{ex.phi}, which was
actually computed in the quadratic approximation~\ref{quad.decomp}
\cite{Bibi97}. Now, it is possible to compute the entropy per tile
associated with $\fm$: a simple numerical calculation gives
$S[\fm]=0.2616(3)$. This value is in exact agreement with the exactly
known diagonal entropy of fixed boundary tilings. These results are
therefore a validation of our continuous approach (coarse graining).
As announced, we have established an exact link between free and fixed
boundary tilings.

\medskip

This function $\fm$ deserves a quick qualitative description; since
$\fm$ is very close to the function displayed in figure~\ref{ex.phi},
we will use this figure to illustrate our arguments: first, when the
boundary is an hexagon, it has a 3-fold symmetry and, as foreseen,
the solution $\fm$ respects this symmetry. Second, because of the
strong influence of the boundary, there is a gradient of entropy
between the boundary and the bulk. Indeed, near the center of the
tiling, the gradient of $\fm$ is very close to the free boundary one,
whereas far from the center, this gradient becomes more and more
influenced by the boundary and becomes singular out of the arctic
circle: there, the entropy is zero. Then the fixed boundary tilings
provide a very interesting model having an inhomogeneous entropy
distribution.

To close this discussion, let us precise that this infinite size limit
cannot be called a ``thermodynamic limit'' because of this lack of
homogeneity: in statistical mechanics, a system is said to be at the
thermodynamic limit if its properties do not depend on how it tends to
infinity. In particular, they must be homogeneous in the system and
must not depend on the container shape (here, the boundary)
\cite{Toda}. Here, the situation is far from this one, since even the
stoichiometry depends on the boundary shape.

\subsection{Other Kinds of Tilings}
\label{Other}

So far, we have only considered zonotopal fixed boundaries. The reason
for this choice is that the first motivation of this work was to
establish the link between free boundary tilings and tilings built by
means of the so-called partition method, the aim of which was to
develop a new approach to these tiling questions
\cite{Elser84,Mosseri93,Mosseri93B,Bibi97,These}. The latter tilings
precisely have zonotopal boundary conditions, by construction.
However, there is no reason why the previous method could not be
applied to other kinds of boundaries. In this section, we will
analyze tilings, the boundary of which is fixed, but which nonetheless
have a free boundary entropy.

Indeed, if the boundary is fixed but imposes a uniform phason
gradient, that is to say the membrane frame lies in a $d$-dimensional
plane of gradient $\vect{E}=\vect{E}_0$, then the affine membrane of
gradient $\vect{E}_0$ has on the one hand the good boundary
conditions, and on the other hand satisfies the partial differential
equation~\ref{edp} (since its second order derivatives are equal to
zero). This membrane is therefore the function $\fm$ and its entropy
$s[\fm]$ is equal to the free boundary one,
$\tau(\vect{E}_0)$.

\medskip

\begin{figure}[ht]
\begin{center}
\deuxfigs{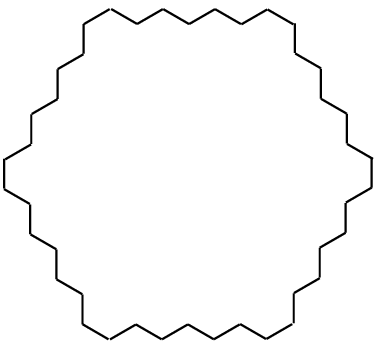}{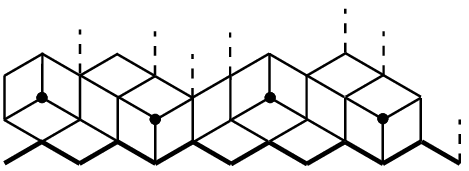}
\end{center}
\caption{Left: a zigzag boundary, without phason strain.
  Right: tiling detail; the boundary (thick line) is globally --
  but not locally -- strait.}
\label{zigzag}
\end{figure}

More precisely, let us analyze a $3 \ra 2$ class of tilings, the fixed
boundary of which is {\sl flat} in the membrane representation in 3
dimensions, that is to say does not impose any global phason strain to
the tilings. This kind of boundary is illustrated in
figure~\ref{zigzag}.

At the infinite size limit and in the membrane point of view, after
rescaling, the boundary becomes a flat hexagon, that is the function
$\phi$ is constrained to zero on this boundary. Then the function $\phi=0$
maximizes the entropy and, at the infinite size limit, this entropy is
equal to $\sigma_0^{free}$.

We have tested this theoretical prediction by a direct calculation of
this entropy. The method is developed in appendix \ref{ex.Flat}. It
uses again the Gessel-Viennot method \cite{Gessel85,Stembridge90}. The
entropy of very large tilings can then be numerically reached and
fitted to get its infinite size limit. We find an entropy per tile
of 0.32306(4), which is precisely the free boundary entropy.

\medskip

To close this section, let us draw attention to numerical simulations
by Joseph and Baake~\cite{Joseph96}: they analyzed the configurational
entropy of random $4 \ra 2$ tilings, the boundary of which is fixed
and flat (in 4 dimensions), as in our previous example (the global
phason strain is zero). As foreseen, the entropy that they eventually
found was equal to the free boundary one, which was itself numerically
estimated.

\section*{Conclusion}

Thanks to a continuous approach in the membrane point of view at the
infinite size limit, we have been able to describe the dominant states
of statistical ensembles of tilings. In particular, this method has
enabled us to establish a quantitative link between two exactly
solvable models of statistical mechanics, the free and fixed boundary
tilings of 60 degree lozenges. In the latter, a very remarkable event
occurs, the arctic circle phenomenon: there exist ``frozen'' regions
of the tilings in which there is only one kind of tiles and were
the entropy is therefore zero. This lack of homogeneity is responsible
for the difference of entropy between the two problems, even if they
were at first sight closely related. Moreover, this infinite size
limit cannot be called a thermodynamic limit because of this lack
of homogeneity.

In the case of larger dimension or codimension tilings, a similar
treatment would require the knowledge of the free boundary entropies.
Unfortunately, despite a great deal of work, these entropies are not
known yet. However, the maximum (diagonal) entropies and the phason
elastic constants are numerically known in many cases. It could
therefore be possible, in these cases, to compute the fixed boundary
entropies and phason elastic constants, in the quadratic
approximation. But this approximate method would not be
conclusive on the existence of an arctic circle phenomenon in such
problems, which is nonetheless a captivating open question.

Finally, it is worth emphasizing that this method could be useful in
describing how any constraint imposed at the boundary relaxes into the
bulk. In this paper, we have studied two kinds of boundaries. The
first one, the straight boundary related to partition problems,
imposes to the tiling the strongest constraint among all boundary
conditions: the tilings must relax continuously from a completely
crystalline structure to a random one. Physically, such tilings (in
3D) could model the result of a growth of quasicrystalline materials
on crystalline phases. More generally, more complex physical
situations, such as extended topological defects (such as
dislocations), or other kinds of interfaces, could {\sl a priori} be
translated in suitable boundary conditions. The numerical method we
have developed could then be applied to any such boundary conditions
and could be useful in describing how the material relaxes in the presence
of such constraints.

\bigskip

\noindent {\sl Note:} we have recently been aware of related works in the
Aztec diamond tiling problem \cite{Cohn9?}.

\appendix

\section{The Gessel-Viennot Method}

In this appendix, we present a combinatorial method for the counting
of configurations of avoiding paths on planar graphs, the
Gessel-Viennot method, which can be very useful in the enumeration of
fixed boundary tilings. We illustrate this method in two $3 \ra 2$
cases discussed in the present paper.

\subsection{The Method}
\label{methode}

We will not extensively present the Gessel-Viennot method
\cite{Gessel85,Stembridge90}. We will instead give a brief description
and try to explicit the underlying ideas. The method is rather
general: it can be applied to any oriented graph without cycles
(acyclic oriented graph), in which are selected two families of
vertices, $u_i$ and $v_i$, $i=1,\ldots,n$. This graph is supposed to
satisfy the property of {\sl compatibility}: if two paths on this
graph are going respectively from $u_{i_1}$ to $v_{j_1}$ and from
$u_{i_2}$ to $v_{j_2}$ and if these paths do not cross, then $i_1 <
i_2$ and $j_1 < j_2$. Note that this property is very specific to
two-dimensional graphs.

We are interested in the number of configurations of $n$ avoiding (or
non-intersecting) paths, the $i$-th path going from $u_i$ to $v_i$. If
we denote by $\lambda_{ij}$ the number of paths going from $u_i$ to
$v_j$, then the method states that the number of configurations is
equal to the following determinant:
\begin{equation}
D_n=\det(\lambda_{ij})_{1 \leq i,j \leq n}.
\end{equation}

The idea of the proof is the following: in this determinant, all path
configurations, whether intersecting or not, the $i$-th path going
from $u_i$ to $v_{\sigma(i)}$, for any permutation $\sigma$, are
counted, with a $+$ or $-$ sign. All intersecting configurations
cancel two by two and only the non-intersecting configurations
remain. They are exactly the sought configurations thanks to the
property of compatibility. The reader interested in more details can
refer to the review paper by Stembridge \cite{Stembridge90}.

\subsection{Hexagonal Boundaries}
\label{ex.Hex}

\begin{figure}[ht]
\begin{center}
\ \psfig{figure=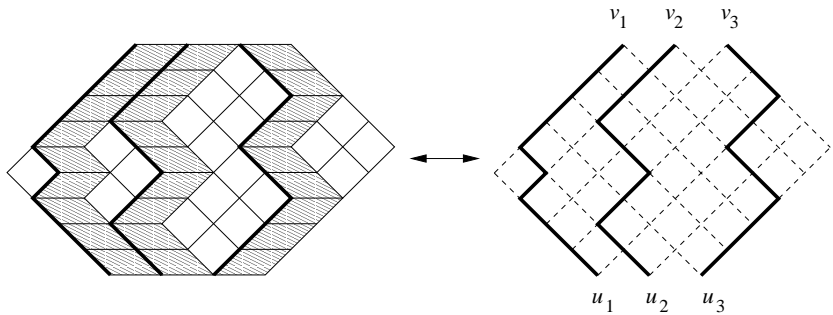,width=11cm} \
\end{center}
\caption{A hexagonal boundary $3 \ra 2$ tiling (left): the dashed worms
  can be translated into a set of $p$ avoiding oriented paths on a square
  lattice (right). The $i$-th path starts from the (fixed) vertex $u_i$
  and goes to the (fixed) vertex $v_i$. There are $p$ paths. The side
  lengths of the hexagon are $p$, $k$ and $l$.}
\label{lignes}
\end{figure}

In this section, the previous method will be used to rederive
MacMahon's formula (see section~\ref{Loz.Til}). Consider a $3 \ra 2$
tiling of an hexagonal region of side lengths $k,l$ and $p$ (figure
\ref{lignes}, left). In such a tiling, one can follow sequences of
tiles which have a horizontal edge. These lines, which are usually
called worms, cross the tiling from bottom to top. The tiling can now
be slightly deformed so that a kind of tiles become squares (figure
\ref{lignes}) and the $p$ worms can be seen as up-going paths on a
square lattice (figure \ref{lignes}, right).

Therefore the previous theory on avoiding paths on an acyclic oriented graph
can be applied. The number $\lambda_{ij}$ of paths joining the
vertices $u_i$ and $v_j$ is a binomial coefficient:
\begin{equation}
\lambda_{ij} = \displaystyle{(k+l)! \over (k+j-i)! \; (l+i-j)!}.
\end{equation}
Then, we have to compute the following determinant:

\begin{equation}
D_p(k,l)= \det (\lambda_{ij}) = \det \left[ (k+l)! \over
(k+j-i)! \; (l+i-j)!
\right]_{1 \leq i,j \leq p}.
\end{equation}

\begin{eqnarray}
D_p(k,l) &=& \left[ (k+l)! \right]^p 
              \det \left[\displaystyle  {1 \over
                (k+j-i)! \; (l+i-j)!}
                \right]_{1 \leq i,j \leq p} \nonumber \\
  & = & \left[ (k+l)! \right]^p \displaystyle{
                {1 \over (k+p-1)! \ldots k!} \ {1 \over (l+p-1)! \ldots l!}} \\
  &  & \times \det \displaystyle{\left[ {(k+p-i)! \over (k+j-i)!} \
                {(l+i-1)! \over (l+i-j)!}
                \right]_{1 \leq i,j \leq p}}. \nonumber
\end{eqnarray}

\bigskip

The first factor equals $\left[ (k+l)! \right]^p
\displaystyle{\fact2{(k-1)} \over \fact2{(k+p-1)}} \ {\fact2{(l-1)}
  \over \fact2{(l+p-1)}}$, where we have used again the second order
factorial function. The second factor is denoted by $\Delta_p$.  We
will use the notation: $\displaystyle{P^{(p)}_j(i)={(k+p-i)! \over
    (k+j-i)!} \ {(l+i-1)! \over (l+i-j)!}}$. Since $j \leq p$,
$P^{(p)}_j$ is a polynomial of degree $(p-j)+(j-1)=p-1$.

We now use the following result concerning polynomials: if $Q_j$,
$j=1,\ldots,p$ are polynomials of degrees smaller than $p-1$, if
$Q_j=\sum\limits_{i=0}^{p-1} a_i^{(j)} X^i$, and if
$x_1,x_2,\ldots,x_p$ are real numbers, then 
\begin{equation}
\det [ Q_j(x_i)
  ]_{1 \leq i,j \leq p} = \det (a_i^{(j)}) \times \VdM
  (x_1,\ldots,x_p),
\label{prod.det}
\end{equation}
where $\VdM (x_1,\ldots,x_p)$ is the Van
der Monde determinant of these real numbers. We recall that 
\begin{equation}
\VdM (x_1,\ldots,x_p) =
\begin{array}{|c c c c c|}
1 & x_1 & x_1^2 & \ldots & x_1^{p-1} \\
1 & x_2 & x_2^2 & \ldots & x_2^{p-1} \\
\vdots & \vdots & \vdots & & \vdots \\
1 & x_p & x_p^2 & \ldots & x_p^{p-1} 
\end{array} \ .
\end{equation}
The proof of equation~\ref{prod.det} is straightforward: the left-hand
side matrix is the product of the coefficient matrix and of the Van
der Monde matrix.  Note that $\VdM(1,2,\ldots,p) = \fact2{(p-1)}$.

\medskip

Now, the calculation of $\Delta_p$ is made by induction on $p$:
if $d_p$ denotes the determinant of the coefficients of the 
polynomials $P^{(p)}_j$, $j=1,\ldots,p$, then thanks to the above results,
$\Delta_{p+1}=\fact2{p} d_{p+1}$. After a tedious calculation, one
gets that
\begin{equation}{d_{p+1} \over d_p} = P_{p+1}^{(p+1)}(k+p+1)=(k+l+p)
  \ldots (k+l+1).
\end{equation}
So by induction on $p$,
\begin{equation}
d_p = {1 \over {\left[ (k+l)! \right]^p}} \ {\fact2{(k+l+p-1)} 
\over \fact2{(k+l-1)}}.
\end{equation}
Finally, we get 
\begin{equation}
D_p(k,l) = {
{\fact2{(k+l+p-1)} \ \fact2{(k-1)} \ \fact2{(l-1)} \ \fact2{(p-1)} }
\over
{\fact2{(k+p-1)} \ \fact2{(l+p-1)} \ \fact2{(k+l-1)}}},
\end{equation}
which is precisely Mac Mahon's enumerative formula \cite{MacMahon}, 
rewritten in terms of generalized factorials.

\subsection{Flat Fixed Boundaries}
\label{ex.Flat}

This method can also be applied to the tilings studied in
section~\ref{Other} (figure~\ref{zigzag}). However, in this case, we
will not be able to get an explicit enumerative formula and some
numerical calculations will be necessary to have access to the
infinite size entropy.

\begin{figure}[ht]
\begin{center}
\ \psfig{figure=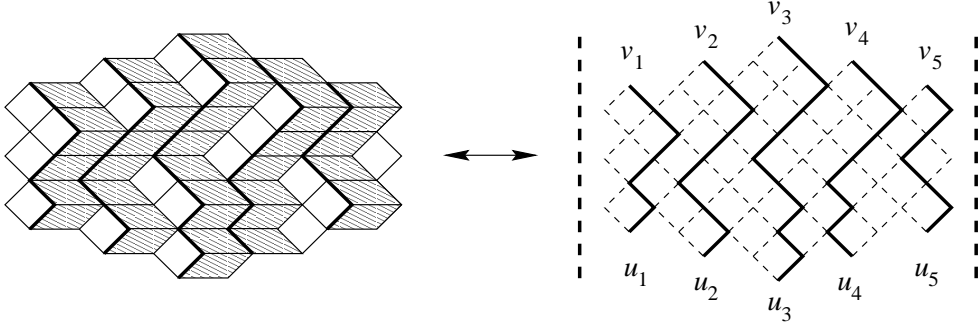,width=13cm} \
\end{center}
\caption{The worm representation of a flat-boundary tiling (left) and its
counterpart in the avoiding path representation (right), in the case when
the number of lines $n$ is equal to 5.}
\label{lignes2}
\end{figure}

With these boundary conditions, the worm and path representations of
figure~\ref{lignes} must be modified, as illustrated in
figure~\ref{lignes2}.

The two main differences are the following: first, the vertices $u_i$
and $v_i$ are not as simply distributed as in figure~\ref{lignes}. Second,
some of the $n$ paths are constrained by the presence of two vertical bounds
(thick dashed lines). Therefore, the number $\lambda_{ij}$ of paths
starting from $u_i$ and going to $u_j$ might be different from a
simple binomial coefficient. This number can be calculated thanks to
the usual ``mirror'' or ``image  method'' (see \cite{Montroll87},
for instance). In the diagonal case, with the indexation of
figure~\ref{lignes2} and when $n$ is odd,
\begin{equation}
\lambda_{ij}=\Simp{p_{ij}}{p_{ij}+3(i-j) \over 2}
-\Simp{p_{ij}}{p_{ij}+3(i+j)-2 \over 2}
-\Simp{p_{ij}}{p_{ij}+6n-3(i+j)+4 \over 2},
\end{equation}
where 
\begin{equation}
p_{ij}=2n - \left|{n+1 \over 2} - j \right|- \left|{n+1 \over 2} - i \right|
\end{equation}
is the length of any path going from $u_i$ to $v_j$.

The method is now strictly similar to the previous one. However, the
complete calculation of the determinant $\det ( \lambda_{ij})$
seems out of reach. That is why we have chosen to compute it
numerically for large systems. The so-obtained values are
displayed in table~\ref{GV2}. In this table, $n$ still denotes the
number of worms.

\begin{table}[ht]
\caption{Entropy per tile of $n$-worm tilings with flat
  boundaries. These boundaries do not impose any global phason strain.}
\label{GV2}
$$
\begin{array}{|c|c|}
\hline
\mbox{Number of worms} & \mbox{Entropy per tile} \\ \hline
n=21 & S=0.311881 \\ 
n=31 & S=0.315379 \\ 
n=61 & S=0.319098 \\
n=81 & S=0.320065 \\
n=101 & S=0.320653 \\ 
n=151 & S=0.321446 \\ \hline
\end{array}
$$
\end{table}
The last four data can be fitted with the following law:
\begin{equation}
S(n)= S_0 - {A \over n} + {B \over n^2}.
\end{equation}
Then we get $S_0=0.32306(4)$,
which is the infinite size entropy (and $A \simeq -0.246$, $B \simeq 0.245$).


\begin{thebibliography}{99}

\bibitem{Shechtman84} D. Shechtman, I. Blech, D. Gratias, J.W. Cahn,
Metallic Phase with Long-Range Orientational Order and No Translational
Symmetry, {\em Phys. Rev. Lett.} {\bf 53}, 1951 (1984).

\bibitem{Penrose74} R. Penrose, The Role of Aesthetics in Pure and
    Applied Mathematical Research, {\em Bull. Inst. Math. Appl.} {\bf
    10}, 226 (1974).

\bibitem{Elser} V. Elser, Comment on ``Quasicrystals: a New Class of
Ordered Structures'', {\em Phys. Rev. Lett.} {\bf 54}, 1730 (1985).

\bibitem{Henley91} C.L. Henley, Random Tiling Models, {\sl in} {\em
    Quasicrystals, the State of the Art}, Ed. D.P.  Di Vincenzo, P.J.
  Steinhart (World Scientific, 1991), 429.

\bibitem{Widom89} M. Widom, D.P. Deng, C.L. Henley, Transfer-Matrix
    Analysis of a Two-Dimensional Quasicrystal, {\em
    Phys. Rev. Lett.} {\bf 63}, 310 (1989).

\bibitem{Strandburg89} K.J. Strandburg, L.H. Tang, M.V. Jari\'c,
    Phason Elasticity in Entropic Quasicrystals, {\em
    Phys. Rev. Lett.} {\bf 63}, 314 (1989).

\bibitem{Elser84} V. Elser, Solution of the Dimer Problem on an Hexagonal 
Lattice with Boundary, {\em J. Phys. A: Math. Gen.} {\bf 17}, 1509 
(1984).

\bibitem{Mosseri93} R. Mosseri, F. Bailly, C. Sire, Configurational 
Entropy in Random Tiling Models, {\em J. Non-Cryst. 
Solids}, {\bf 153}\&{\bf 154}, 201 (1993).

\bibitem{Mosseri93B} R. Mosseri, F. Bailly, Configurational Entropy
in Octagonal Tiling Models, {\em Int. J. Mod. Phys. B}, Vol 7, {\bf 6}\&{\bf
7}, 1427 (1993).

\bibitem{Bibi97} N. Destainville, R. Mosseri, F. Bailly,
Configurational Entropy of Codimension-One Tilings and Directed Membranes, 
{\em J. Stat. Phys.} {\bf 87}, Nos 3/4, 697 (1997).

\bibitem{These} N. Destainville, Ph.D. Thesis: ``Entropie
  configurationnelle des pavages aléatoires et des membranes
  dirigées'', Thèse de l'Université Paris 6 (1997).

\bibitem{Wannier50} G.H. Wannier, Anti-ferromagnetism. The Triangular Ising 
Net, {\em Phys. Rev.} {\bf 79}, 357 (1950); {\em Phys. Rev.} B {\bf 7} 5017
(E, 1973).

\bibitem{Bibi95} N. Destainville, R. Mosseri, F. Bailly,
Role of Boundary Conditions in Configurational Entropy of Random
Tilings, {\sl in} {\em Proceedings of the 6th International Conference on
Quasicrystals} (World Scientific, 1997).

\bibitem{Duneau} M. Duneau, A. Katz, Quasiperiodic Patterns, {\em
    Phys. Rev. Lett.} {\bf 54}, 2688 (1985).

\bibitem{Kalugin} A.P. Kalugin, A.Y. Kitaev, L.S. Levitov, 
Al$_{0.86}$Mn$_{0.14}$: a Six-Dimensional Crystal, {\em JETP Lett.} {\bf
41}, 145 (1985); A.P. Kalugin, A.Y. Kitaev, L.S. Levitov, 6-Dimensional
Properties of Al$_{0.86}$Mn$_{0.14}$, {\em J. Phys. Lett. France} {\bf 46},
L601 (1985).

\bibitem{Coxeter} {\em Regular Polytopes}, H.S.M. Coxeter (Dover,
  1973).

\bibitem{MacMahon} {\em Combinatory Analysis}, P.A. Mac Mahon (Cambridge
University Press, 1916).

\bibitem{Gessel85} I. Gessel, G. Viennot, Binomial Determinants, Paths,
and Hook Length Formulae, {\em Advances in Mathematics} {\bf 58}, 300--321
(1985).

\bibitem{Stembridge90} J.M. Stembridge, Non-Intersecting Paths, Pfaffians,
and Plane Partitions, {\em Advances in Mathematics}
{\bf 83}, 96--131 (1990).

\bibitem{Toda} {\em Statistical Physics I}, M. Toda, R. Kubo,
  N. Saitô (Springer-Verlag, Berlin, 1982).
  
\bibitem{Blote82} H.W.J. Blöte, H.J. Hilhorst, 
Roughening Transitions and Zero-Temperature Triangular Ising
Anti-Ferromagnet, {\em J. Phys. A: Math.  Gen.} {\bf 15}, L631 (1982).

\bibitem{Kalugin94} P.A. Kalugin, The Square-Triangle Random-Tiling
  Model in the Thermodynamic Limit, {\em J. Phys. A: Math. Gen.} {\bf
  27}, 3599 (1994).

\bibitem{Nienhuis96} J. de Gier, B. Nienhuis, Exact Solution of an
  Octagonal Random Tiling Model, {\em Phys. Rev. Lett.} {\bf 76}, 2918
  (1996).

\bibitem{Houtappel50} R.M.F. Houtappel, {\em Physica} {\bf 16}, 425 (1950).

\bibitem{Cohn98} H. Cohn, M. Larsen, J. Propp, The Shape of a Typical
  Boxed Plane Partition, {\sl submitted to} {\em New York J. of
  Math.}

\bibitem{Propp98} J. Propp, Boundary-Dependent Local Behavior for 2-D
  Dimer Models, {\sl to appear in} {\em International J. of Mod. Phys.}

\bibitem{Joseph96} D. Joseph, M. Baake, Boundary Conditions, Entropy,
    and the Signature of Random Tilings, {\em J. Phys. A:
    Math. Gen.} {\bf 29}, 6709 (1996).

\bibitem{Cohn9?} H. Cohn, R. Kenyon, J. Propp, A Variational Principle
  for Domino Tilings, {\sl preprint}.

\bibitem{Montroll87} E.W. Montroll, B.C.J. West, On an Enriched
  Collection of Stochastic Processes, {\sl in} {\em Fluctuation
    Phenomena}, Ed. E.W. Montroll, J.L. Lebowitz, (North-Holland, 1987).


\end{thebibliography}
\end{document}